\documentclass[a4paper,11pt]{article}
\pdfoutput=1 

\usepackage[T1]{fontenc} 
\usepackage{lineno}
\usepackage{url}

\usepackage{blindtext}
\usepackage{amsmath}
\usepackage{amssymb}
\usepackage{epsfig}
\usepackage{graphicx}
\usepackage{amsmath}
\usepackage{geometry}
\usepackage{graphicx}
\usepackage{microtype}
\usepackage{multirow}
\usepackage{float}
\usepackage{rotating}
\usepackage[section]{placeins}
\usepackage{anysize}
\usepackage{enumerate}
\usepackage{subcaption}

\usepackage{color}

\usepackage{url}
\usepackage{soul}

\DeclareFixedFont\trfont{OT1}{phv}{b}{sc}{11}
\newif\ifnotoc\notocfalse
\newif\ifemailadd\emailaddfalse
\newif\iftoccontinuous\toccontinuousfalse
\newif\ifnatbibsort\natbibsorttrue

\relax

\RequirePackage{amsmath}
\RequirePackage{amssymb}
\RequirePackage{epsfig}
\RequirePackage{graphicx}
\ifnatbibsort\RequirePackage[numbers,sort&compress]{natbib}\else\RequirePackage[numbers,compress]{natbib}\fi
\RequirePackage[colorlinks=true
  ,urlcolor=blue
  ,anchorcolor=blue
  ,citecolor=blue
  ,filecolor=blue
  ,linkcolor=blue
  ,menucolor=blue
  ,pagecolor=blue
  ,linktocpage=true
  ,pdfproducer=medialab
  ,pdfa=true
]{hyperref}

\numberwithin{equation}{section}
\def\fnum@figure{\textbf{\figurename\nobreakspace\thefigure}}
\def\fnum@table{\textbf{\tablename\nobreakspace\thetable}}

\renewenvironment{thebibliography}[1]{%
\begin{oldthebibliography}{#1}%
\small%
\raggedright%
\setlength{\itemsep}{5pt plus 0.2ex minus 0.05ex}%
}%
{%
\end{oldthebibliography}%
}

\usepackage{makeidx}  
\usepackage{graphicx} 
\usepackage{booktabs} 
\usepackage{url}
\usepackage{lipsum}
\usepackage{amsmath}
\usepackage{mathtools}

\usepackage{hyperref}
\usepackage{graphicx}
\usepackage{xspace}

\providecommand{\PYTHIA} {{\textsc{pythia}}\xspace}

\providecommand{\PROFESSOR} {\textsc{professor}\xspace}
\providecommand{\RIVET} {\textsc{RIVET}\xspace}

\newcommand{\alpS}{\ensuremath{\alpha_{s}}\xspace}

\newcommand{\rpv}{\ensuremath{\rlap{\kern.2em/}R}\xspace}

\usepackage{url}
\usepackage{soul}

\newcommand{\PY}        {\textsc{pythia8}\xspace}

\newcommand{\cindf}     {\ensuremath{\chi^{2}/NDF}\xspace}  

\newcommand{\taup}      {\tau_{\perp}}
\newcommand{\mass}      {\rho_{\text{Tot}}}
\newcommand{\massp}     {\rho^{\textrm{T}}_{\text{Tot}}}
\newcommand{\bT}        {B_\textrm{T}}
\newcommand{\HT}        {H_{\textrm{T,2}}}
\newcommand{\ak}        {\textrm{anti-}k_{\textrm{T}}}
\newcommand{\Pevol}     {p_{\perp}}

\begin{document}
\title{\bf Investigating the parton shower model in \textsc{pythia8} with pp collision data at \boldmath{ $\mathrm{\surd{s}=13\, TeV}$}
\thanks{Presented at XXIV DAE-BRNS High Energy Physics Symposium}}
\author{S. K. Kundu$^1$\thanks{sumankundu.rs@visva-bharati.ac.in},    T. Sarkar$^2$\thanks{tanmay.sarkar@cern.ch},   M. Maity$^3$\thanks{manas.maity@visva-bharati.ac.in}  }

\date{%
	$^{1,3}$ Visva-Bharati University, Santiniketan, India\\%
	$^2$ National Central University ( NCU), Taiwan.\\
}

\maketitle              

\begin{abstract}
Understanding the production of quarks and gluons in high energy collisions and their evolution is a very active
area of investigation. Monte carlo event generator \textsc{pythia8} uses the parton shower model  to simulate such
collisions and is optimized using experimental observations. Recent measurements of event shape variables and
differential jet cross-sections in pp collisions at $\mathrm{\surd{s} = 13\, TeV}$ at the Large Hadron Collider
have been used to investigate further the parton shower model as used in \textsc{pythia8}.

\end{abstract}

\section{Introduction}\label{sec:1}

Matrix element calculations with fixed order treatment is not sufficient to understand the production of quarks and gluons, 
collectively called partons, in high energy collisions or their evolution into jets of hadrons. Comparison with experimental 
results demand fully exclusive description of the final states based on the shower evolution and hadronization. 
Such methods are described through phenomenological models embedded in the shower Monte Carlo (MC) codes.
  
  \PY uses leading order(LO) calculations followed by `transverse momentum' ($p_{\perp}$)  ordered parton shower\cite{Sjostrand:2004ef} 
  with  $p^{2}_{\perp}$ as evolution variable for the generation of $2\rightarrow n$ ($n\ge 2$) final states by taking account
  initial (ISR) and final (FSR) state shower. Shower evolution for a parton like $a\rightarrow bc$, is based on the standard (LO) DGLAP splitting kernels and the branching probability expressed as:
\begin{equation}
d\mathcal{P}_{a}=\frac{dp^{2}_{\perp}}{p^{2}_{\perp}}\sum_{b,c}\frac{\alpha_{s}(p^{2}_{\perp})}{2\pi}P_{a\rightarrow bc}(z)dz 
\label{DGLAP}
\end{equation}
where $\mathrm{P_{a\rightarrow bc}}$ is the DGLAP splitting function and $p^{2}_{\perp}$ represents the
scale of the branching; $z$ represents the sharing of $\Pevol$ of $a$ between the
two daughters, with $b$ taking a fraction $z$ and $c$ the rest, $1-z$. Here the summation goes over
all allowed branchings, e.g. $q\rightarrow qg$ and $q\rightarrow q\gamma$ and etc. Now,
the divergence at $p^{2}_{\perp}\rightarrow 0$ is taken care of by introducing a term $\mathcal{P}^{no}_a(p^{2}_{{\perp}_{\textrm{max}}},p^{2}_{{\perp}_{\textrm{evol}}})$
known as \emph{Sudakov form factor} \cite{Sudakov:1954sw}. This Sudakov factor ensures that there will 
be no emission between scale $p^2_{{\perp}_{\textrm{max}}}$ to a given $p^2_{{\perp}_{\textrm{evol}}}$.

Considering lightcone kinematics, evolution variables $\mathrm{p^{2}_{{\perp}_{\textrm{evol}}}}$ for $a\rightarrow bc$ 
at virtuality scale $Q^2$ for space-like branching (ISR) and time-like branching (FSR) are given by $(1-z)Q^2 $  
and $z(1-z)Q^2$ respectively. Finally, equations \ref{ISR_evol} and \ref{FSR_evol} describe the evolutions for ISR and 
FSR respectively \cite{Sjostrand:2004ef}.
\begin{eqnarray}
\textrm{d}\mathcal{P}_{b} & = & \frac{\textrm{d}p^{2}_{{\perp}_{\textrm{evol}}}}{p^{2}_{{\perp}_{\textrm{evol}}}}\frac{\alpha_s(p^{2}_{{\perp}_{\textrm{evol}}})}{2\pi}
        \frac{x^\prime f_a (x^\prime , p^{2}_{{\perp}_{\textrm{evol}}})}{xf_a (x, p^{2}_{{\perp}_{\textrm{evol}}})}
        P_{a\rightarrow bc}(z) dz \mathcal{P}^{\textrm{no}}_b (x,p^{2}_{{\perp}_{\textrm{max}}},p^{2}_{{\perp}_{\textrm{evol}}}) \label{ISR_evol} \\
\textrm{d}\mathcal{P}_{a} & = & \frac{\textrm{d}p^{2}_{{\perp}_{\textrm{evol}}}}{p^{2}_{{\perp}_{\textrm{evol}}}}\frac{\alpha_s(p^{2}_{{\perp }_{\textrm{evol}}})}{2\pi}P_{a\rightarrow bc}(z) \textrm{d}z \mathcal{P}^{\textrm{no}}_a (p^{2}_{{\perp}_{\textrm{max}}},p^{2}_{{\perp}_{\textrm{evol}}})  \label{FSR_evol}
\end{eqnarray}

Currently both the running re-normalisation and factorisation shower scales, i.e. the scales at
which  $\alpS$ and the PDFs are evaluated, are chosen to be $p^2_{\perp_\textrm{evol}}$ \cite{Corke:2010yf}. 
The general methodology of \PY for ISR, FSR and MPI is to start from some maximum scale $p^{2}_{{\perp}_{\textrm{max}}}$
and evolve downward in energy towards next branching untill the daughter partons reach some cut-off.

\begin{table}[tbp]
\centering
\setlength\tabcolsep{11pt}
\begin{tabular}{|cccc|}
\hline   \PY &   Monash & Sampling range &  Optimized  \\
                Parameters set & values &   & values   \\
        \hline {\tt SpaceShower:alphaSvalue}& 0.1365 & $0.1092-0.1638$ & $0.11409^{+0.00078}_{-0.00073}$  \\
           {\tt TimeShower:alphaSvalue}& 0.1365 & $0.1092-0.1638 $ &  $0.15052^{+0.00077}_{-0.00076}$  \\
           {\tt SpaceShower:PTmaxFudge}& 1.0 & $0.6-1.4$ &  $0.9323^{+0.0065}_{-0.0064}$  \\
           \hline
\end{tabular}
\caption{Optimized result of three parameters of \PY is shown along with their default values in the Monash
tune and the sampling range.}
\label{tab:Param1}
\end{table}

\section{Optimizing the Parton Shower Model of PYTHIA8}\label{sec:2}
CMS and ATLAS have done several tunning of \PY around its Monash tune \cite{Skands:2014pea} for underlying event (UE), 
the strong coupling, and MPI related parameters \cite{Khachatryan:2015pea,Sirunyan:2019dfx} \cite{Buckley:2014ctn}. In this 
study\cite{Kundu:2019scu} Monash tune also used as default for \PYTHIA v8.235 with NNPDF2.3 PDF (LO) set to optimize with 
four event shapes \cite{Sirunyan:2018adt} measurement from CMS. These are - the complement of transverse thrust ($\mathrm{\taup}$), 
total jet mass ($\mathrm{\mass}$), total transverse jet mass ($\mathrm{\massp}$) and total jet broadening ($\mathrm{\bT}$). 

  Monash tune overestimates the multijet regions of these event shapes \cite{Sirunyan:2018adt}, Hence we examined 
ISR and FSR utilising the provision that \PY allows the use of separate values of $\mathrm{\alpS(M_{Z})}$ for the
showering frameworks used for these. The maximum evolution scale involved in the showering is set to match the scale of the hard process 
itself. In \PY it is set equal to the factorization scale, but allows its modification by multiplicative factors 
{\tt SpaceShower:PTmaxFudge} for ISR and {\tt TimeShower:PTmaxFudge} for FSR. The latter is seen not to have much 
effect on the ESVs, it is excluded from the optimization.

For each point in the parameter space, resulting distributions have been compared with data in terms of $\mathrm{\cindf}$. 
Then \PROFESSOR v2.3.0 \cite{Buckley:2009bj} along with \RIVET v2.6 \cite{Buckley:2010ar} has been used to optimize the 
complete set of ESV distributions from \PY \cite{Sirunyan:2018adt}. Post optimization, the new parameter 
set is checked\cite{Kundu:2019scu} using other relevant results from the CMS \cite{Khachatryan:2016wdh} and ATLAS \cite{ATLAS-CONF-2016-092}.

\begin{figure}[tbp]
\centering
  \includegraphics[page=3, width=.40\textwidth, height =.40\textwidth ]{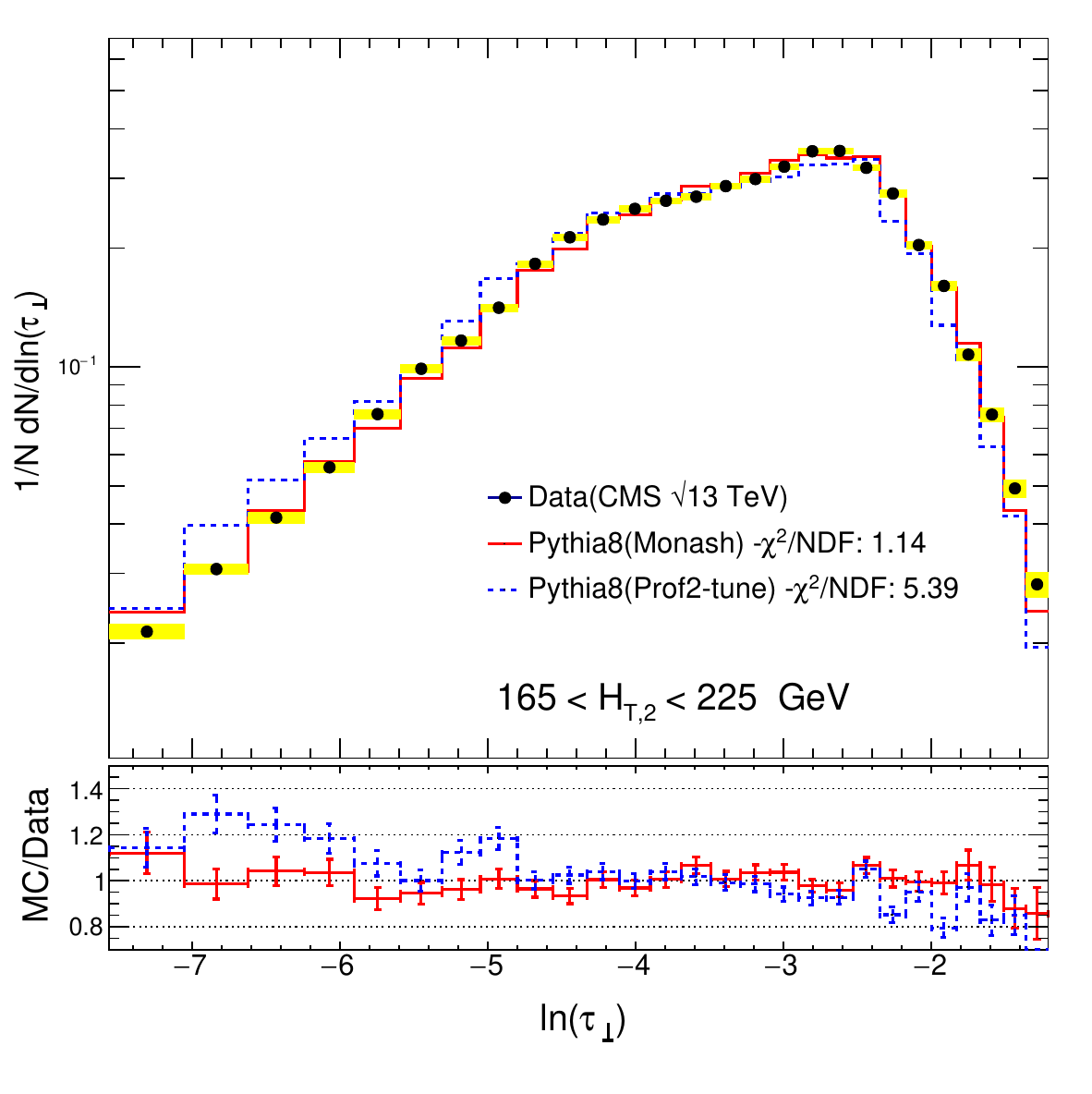}
  \includegraphics[page=11, width=.40\textwidth, height =.40\textwidth ]{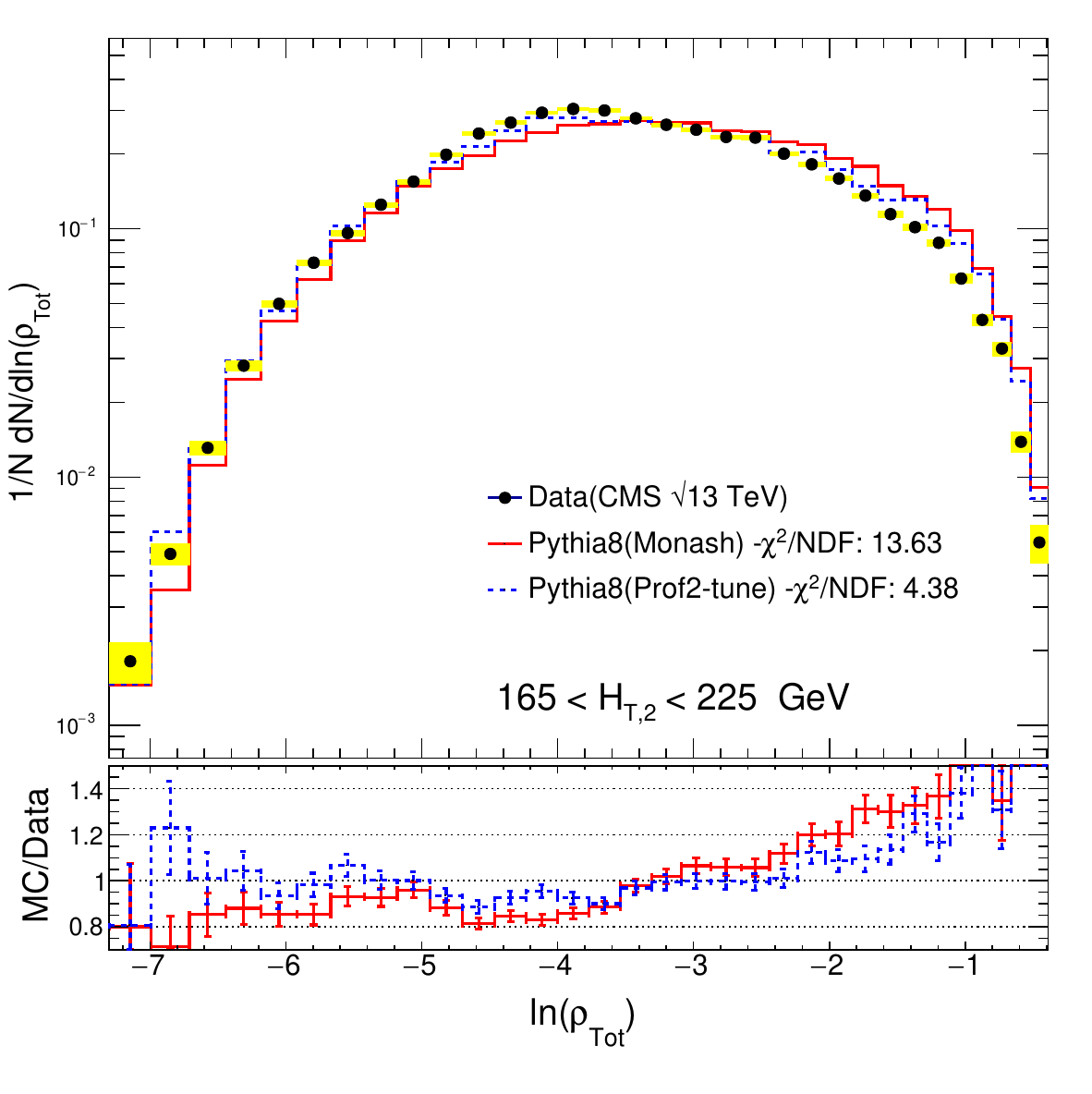}
  \includegraphics[page=19, width=.40\textwidth, height =.40\textwidth ]{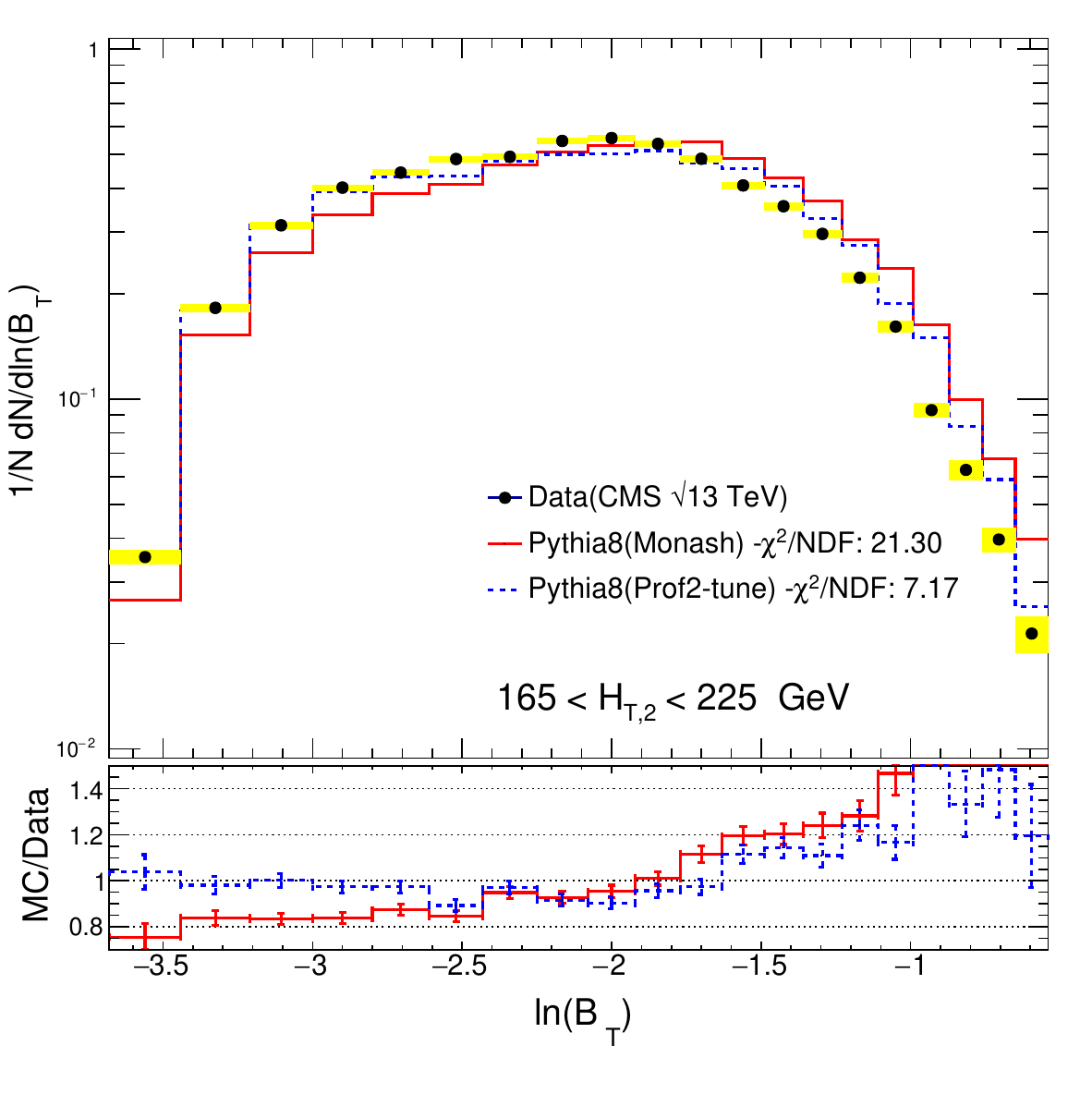}
  \includegraphics[page=27, width=.40\textwidth, height =.40\textwidth ]{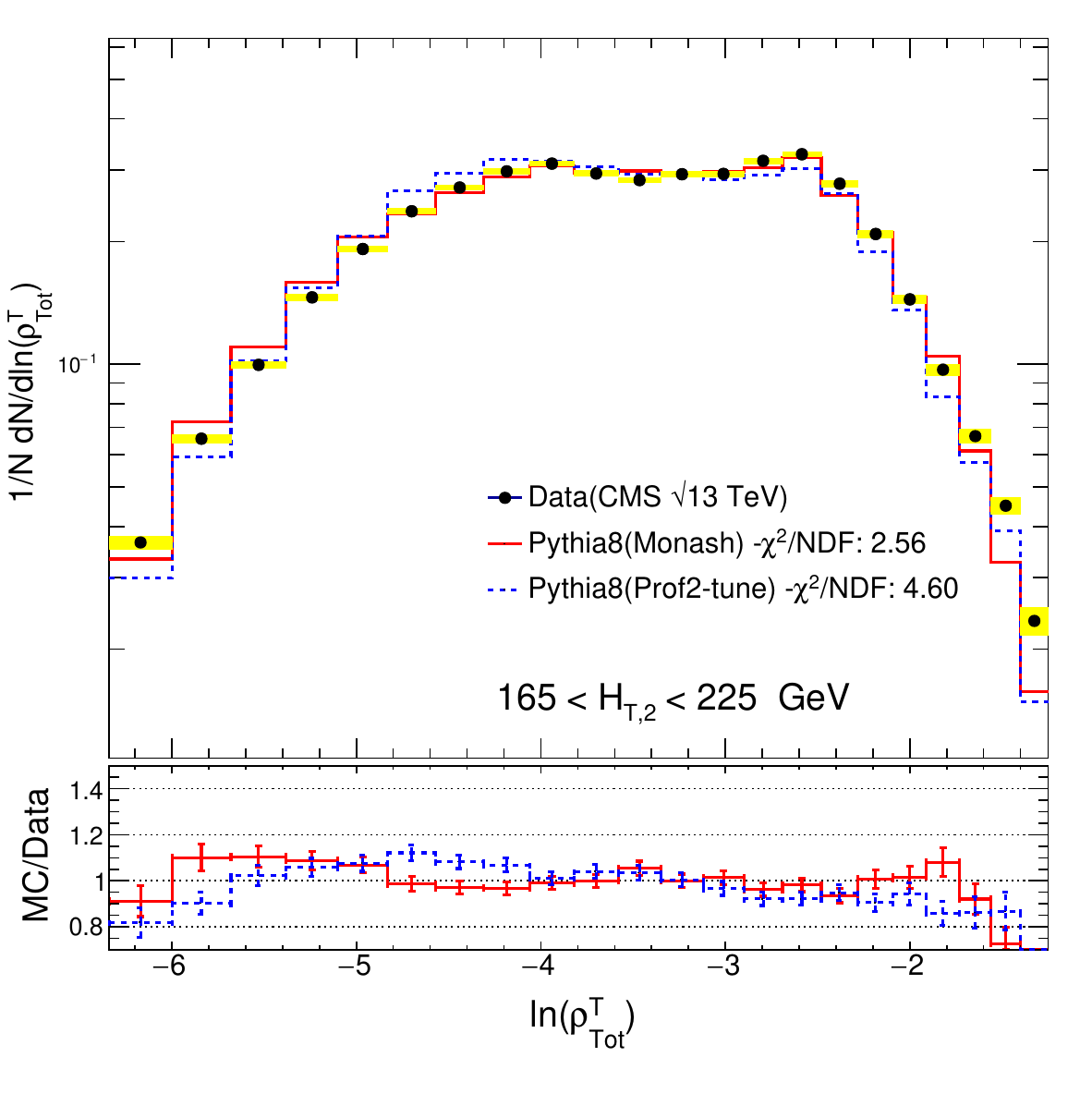}

\caption{Predictions of the optimized parameter set is compared with CMS data and Monash tune for $\HT$ range $165< \HT < 225 $. 
	normalized distributions of the $\mathrm{\taup}$(top left), $\mathrm{\mass}$(top right), $\mathrm{\bT}$(bottom left) and $\mathrm{\massp}$(bottom right)}
\label{Fig:Monash_taup}
\end{figure}

\begin{figure}[tbp]
\centering
\includegraphics[page=9, width=.40\textwidth, height =.40\textwidth ]{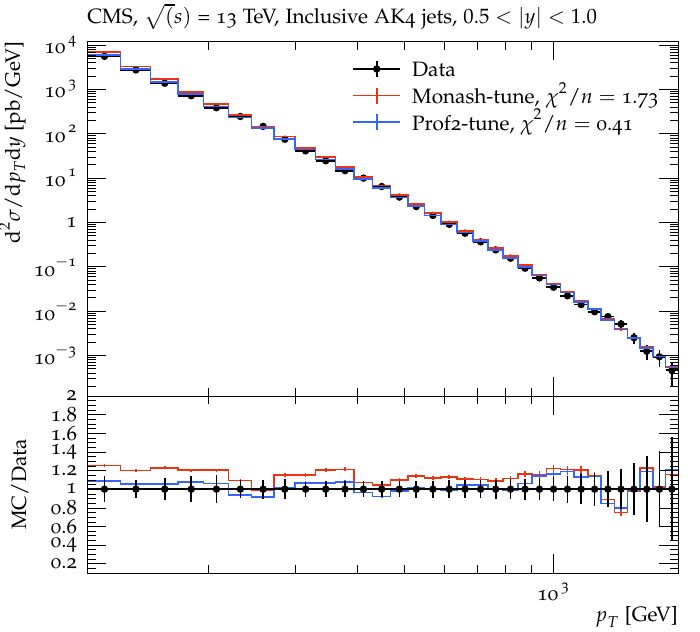}
\includegraphics[page=2, width=.40\textwidth, height =.40\textwidth ]{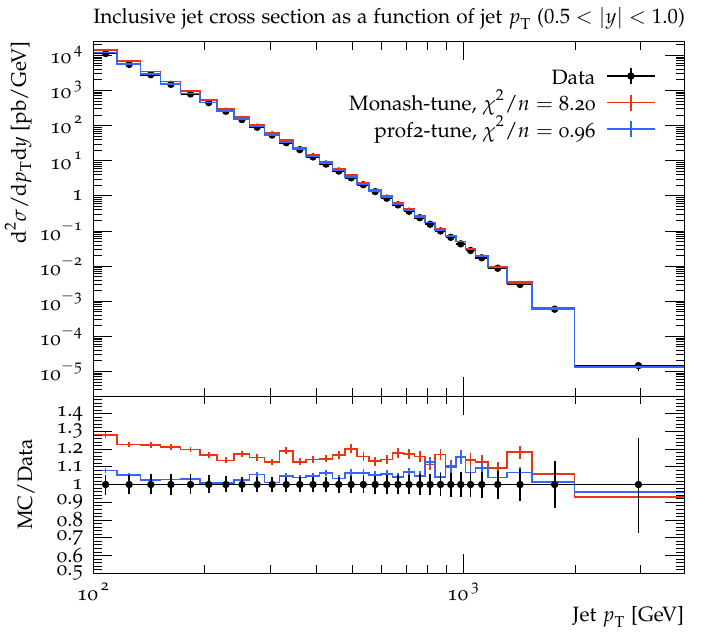}
\caption{Normalized distributions of differential inclusive cross-section for $\ak$ jets (R=0.4) for CMS(left) and ATLAS(right) are
	compared with the predictions of \PY with the optimized parameter set and Monash tune.}
\label{Fig:CMSinc_jet1}

\end{figure}

\section{Validation of results}
The optimized values of the three parameters (see, table \ref{tab:Param1}) are used to calculate the ESVs.
Agreement with data deteriorates slightly for $\mathrm{\taup}$ and $\mathrm{\massp}$ 
(figure \ref{Fig:Monash_taup}) compared to the good agreement with the Monash tune. But, there is significant 
improvement in agreement with data for $\mathrm\mass$ and $\mathrm{\bT}$ (figure \ref{Fig:Monash_taup})compared 
to the Monash tune. Since $\mathrm\mass$ and $\mathrm{\bT}$ had a rather poor agreement between data and the 
Monash tune, overall this new set of parameters is better.

Inclusive jet cross-section measurements being sensitive to PDF of protons and 
$\mathrm{\alpS}$ are also compared with those optimized values. CMS~\cite{Khachatryan:2016wdh} and ATLAS~\cite{ATLAS-CONF-2016-092} studies 
with the 13 TeV data considered for this validation. The CMS measurements of inclusive cross-sections for $\ak$ jets with 
R = 0.4, 0.7. Figures \ref{Fig:CMSinc_jet1} show that the new parameter set improves the agreement 
between data and the Monash tune of Pythia8. Similar improvement is seen for the ATLAS measurement of $\ak$ 
jets with R = 0.4 (figure \ref{Fig:CMSinc_jet1}). 

Since \PY is widely used, its optimization is important. This study 
shows that certain aspects of the experimental observations can be better described with this optimized set of parameters.

\bibliographystyle{spphys}
\bibliography{SMP-17-003}
	

\end{document}